\documentclass[aps, prb,twocolumn, superscriptaddress, showpacs, preprintnumbers,amsmath,amssymb]{revtex4-2}

\usepackage{amsmath}
\usepackage[T2A]{fontenc}
\usepackage[utf8]{inputenc}
\usepackage[english,russian]{babel}
\usepackage{graphicx}
\usepackage{wrapfig}

\usepackage{bm}

\DeclareMathSizes{10}{9}{7}{5}
\DeclareMathOperator{\sign}{sign}

\date{}
\DeclareMathOperator{\Tr}{Tr}

\begin{document}
\selectlanguage{english}

\title{Non-stationary transport properties of boundary states in Kitaev chain}

\author{Yu.~M.~Bilinsky}
\author{P.~I.~Arseyev}
\email{ars@lpi.ru}
\affiliation{ P.N. Lebedev Physical Institute of RAS, 119991 Moscow, Russia}
\author{N.~S.~Maslova}
\affiliation{Chair of Quantum electronics and Quantum Technology
Center, Physics Department, Lomonosov Moscow State University,
119991 Moscow, Russia}

\date{\today}

\begin{abstract}
We investigate the role of gap states in processes of perturbation transmission along finite superconducting Kitaev chain.  We look at this problem on the general ground and use the formalism of non-stationary Green's functions, which contain full information about the non-equilibrium and non-stationary properties of the system.
We discuss tunneling current and non-stationary transport properties of a finite Kitaev chain with each edge connected to its own external lead.
It is shown that the tunneling current is always exponentially small for long chains.
The time-dependent behavior of the tunneling current after the sudden change of bias voltage in one of the leads is also obtained. We investigate the characteristic time of charge transfer from the state at one end of the chain to the opposite edge state. We obtain that this time always exponentially increases with the growth of the chain length, and the relaxation time to the new equilibrium occupation number for the localized state is very large.
Our calculations are completely analytical and straightforward, in contrast with many other methods.
Obtained results show how quickly the \,''second half'' of Majorana state responds after external perturbation acts on the \,''first half''
and why "Majorana" states can hardly be used for any practical devices that require signal transmission from one end of the system to the other.
\end{abstract}

\pacs{74.20.Fg, 74.78.-w, 74.25.-q}
\maketitle

\section{Introduction}

For any practical use, we have to be able to estimate some definite physical characteristics for a system under consideration in addition to an interesting theoretical interpretation of the ground state formation or equilibrium spectrum properties of the system.
One of the simplest models which demonstrate spectral properties that allow topological interpretation is an atomic chain with p-wave superconductivity of spinless particles, proposed by A.Kitaev \cite{kit}.
Possible experimental realizations of this model are usually based on the proximity effect in semiconducting wires with strong spin-orbit interaction placed on a superconductor substrate \cite{Exp1,Exp2,Exp3}.
The main interest in the model was caused by the topological interpretation of its ground state properties.
 It was shown that due to ``topological reasons'' there are quantum states localized around the edges of the chain within the superconducting gap.
   These states, often called  ``Majorana states'', are usually associated with the existence of quasiparticles \cite{valk}, associated with Majorana fermions \cite{maj}. Some studies suggest that they could be used as an error-proof way to store and transmit information in quantum technology \cite{QI1,QI2}. However, if a state is protected from arbitrary changes due to external noise, the same protection could make it equally difficult to change the state of the system purposefully as well, which in turn could make it impossible to use the system for any practical applications.
   One possible way to explore how well the system reacts to the signal is to explore its non-stationary transport properties.

To investigate the role of these localized states in the non-stationary transport properties of Kitaev chain we will use the formalism of non-stationary Green's functions.

It will be shown below that this approach allows to obtain analytically explicit expressions for the tunneling current contrary to more complicated methods with ill controlled approximations discussed for example in \cite{MasterEq}.

The exact electron Green's functions for the infinite Kitaev chain in the equilibrium can be found analytically \cite{arrachea}. They can be used to find the non-stationary Green's functions of a finite chain, which allow us to see how the system develops in time if we apply some perturbation to it.
  The idea of our consideration is to treat a finite Kitaev chain as a cut infinite chain, or a chain with strong defects (for a chain with one cut see e.g. \cite{zazunov}). This trick allows us to use Green's functions of an infinite chain to explore all single-particle states in the system.
Our calculations do not require any interpretation of singularities in single-particle Green function as some special "states".
Let us note here that poles of single-particle Green's functions that appear in the gap of the superconductor in this model can hardly be interpreted as single-particle excitations. The genuine Majorana particles discussed in the pioneer papers \cite{maj} are well-defined particles (quasiparticles) with the ordinary algebra of creation and annihilation operators. In any physical problem, such real particles contribute to a single-particle Green's function with residue equal to unity.
It is known that ordinary superconductors with defects, such as paramagnetic \cite{rusinov} or resonance \cite{volk} impurities, often have bound states localized around these defects with energies lying inside the gap. Such states are true single-particle states. In the problem under consideration, we see that the appearance of the poles in the electron Green function in the gap with residues equal to $1/2$ is more an artifact of the model which has a degenerate (in highly symmetrical case) ground state than the appearance of new quasiparticles.

In the present paper, we apply an approach that is based on non-equilibrium single-particle Green's functions of a finite Kitaev chain. In the first step, explicit analytical expressions for these Green functions are derived. Then the time-dependent tunneling current is calculated using obtained Green functions.
It was shown that tunneling current through "Majorana" gap states is always exponentially small for long chains.
We also investigate the characteristic time of charge transfer from the state at one end of the chain to the opposite edge state. We found out that relaxation time to the new equilibrium occupation number of the \,''second half'' of Majorana state after external perturbation acts on the \,''first half'' is very large and exponentially increases with the growth of the chain length.

\section{Green's function approach to a finite Kitaev chain}

The model Hamiltonian of the Kitaev chain can be written as
\begin{equation}
\begin{aligned}
&\hat {\tilde H} =
	- \mu \sum_{n = 1}^{N} \psi^{\dagger}_{n} \psi_{n}
	- t \sum_{n = 1}^{N-1} \left( \psi^{\dagger}_{n} \psi_{n+1} + \psi^{\dagger}_{n+1} \psi_{n} \right)
\\&
	+ \sum_{n = 1}^{N-1} \left( \Delta \psi^{\dagger}_{n} \psi^{\dagger}_{n+1} + \Delta^{*} \psi_{n+1} \psi_{n} \right).
\end{aligned}
\label{ham_zero_000}
\end{equation}

Here $\psi^{\dagger}_{n}$ and $\psi_{n}$ are the creation and annihilation operators of an electron at site $n$; $\mu$ is the chemical potential; $t$ is the hopping parameter between two neighboring sites; $\Delta$ is the superconducting order parameter, which is considered an external parameter in this model; $N$ is the number of sites in the chain.
To find the exact solutions for Green's functions for this Hamiltonian, it is convenient to use Green's functions of the infinite Kitaev chain. Indeed, it is possible to emulate the behavior of a finite chain by using the infinite chain with infinitely strong point defects \nobreak{$U\rightarrow +\infty$} added at sites $0$ and $N+1$. As a result, the particles located between these two sites will be completely isolated from the external parts of the chain, and Green's functions will be identical to the ones of a finite chain of length $N$ (if the site number of the argument lies between $0$ and $N+1$). Thus, the behavior of the system described by the Hamiltonian
\begin{equation}
\hat H = \hat H_0 + \hat V,
\label{ham_orig_short}
\end{equation}
where
\begin{align*}
\hat H_0 =&
	- \mu \sum_{n} \psi^{\dagger}_{n} \psi_{n}
	- t \sum_{n} \left( \psi^{\dagger}_{n} \psi_{n+1} + \psi^{\dagger}_{n+1} \psi_{n} \right)
\\&
	+ \sum_{n} \left( \Delta \psi^{\dagger}_{n} \psi^{\dagger}_{n+1} + \Delta^{*} \psi_{n+1} \psi_{n} \right),
\\
\hat V =&~ U \left( \psi^{\dagger}_{0} \psi_{0} + \psi^{\dagger}_{N+1} \psi_{N+1} \right),
\end{align*}
is equivalent to the one corresponding to Hamiltonian (\ref{ham_zero_000}).

To find the properties of the states in the chain, we will use the formalism of normal and anomalous retarded Green's functions, denoted as \nobreak{$G^{R}_{nm} (t, t')$, $F^{R}_{nm} (t, t')$,} respectively. Hereafter we will use the following definitions:

\begin{widetext}
\begin{equation}
\Gamma^{R}_{nm} (t, t') =
	\begin{pmatrix}
		G^{R}_{nm} (t, t')	&	F^{R}_{nm} (t, t')
		\\
		F^{R\dagger}_{nm} (t, t')	&	G^{R\dagger}_{nm} (t, t')
	\end{pmatrix}
=
	- i
	\left<
		\begin{pmatrix}
			\left\{\psi_{n} (t), \psi^{\dagger}_{m} (t') \right\}	&	\left\{\psi_{n} (t), \psi_{m} (t') \right\}
			\\
			\left\{\psi^{\dagger}_{n} (t), \psi^{\dagger}_{m} (t') \right\}	& \left\{\psi^{\dagger}_{n} (t), \psi_{m} (t') \right\}	
		\end{pmatrix}
	\right>
	\theta(t - t').
\label{gamma_194}
\end{equation}
\end{widetext}
Here $\left\{\hat a, \hat b\right\} = \hat a \hat b + \hat b \hat a$, $ \left< \hat a \right> = \Tr (\hat \rho \hat a )$. As shown in Ref. \cite{arrachea}, the exact solution for Green's functions of the infinite chain can be found as
\begin{equation}
\begin{aligned}
&\Gamma^{0R}_{nm} (\omega)  =
- \frac
	{1}
	{4 (\left|\Delta\right|^2 - t^2) (A_{+}-A_{-})}
\\ \times&
 \left[
	\chi_{+}^{\left|n-m\right|}
\begin{pmatrix}
	\frac
		{\omega - \mu - 2 t A_{+}}
		{\sqrt{A_{+}^2- 1} }
&
	2\Delta \sign \left( n - m \right)
\\
	- 2\Delta^{*} \sign \left( n - m \right)
&
	\frac
		{\omega + \mu + 2 t A_{+}}
		{\sqrt{A_{+}^2- 1} }
\end{pmatrix}
\right.-\\-&\left.
	\chi_{-}^{\left|n-m\right|}
\begin{pmatrix}
	\frac
		{\omega - \mu - 2 t A_{-}}
		{\sqrt{A_{-}^2- 1} }
&
	2\Delta \sign \left( n - m \right)
\\
	- 2\Delta^{*} \sign \left( n - m \right)
&
	\frac
		{\omega + \mu + 2 t A_{-}}
		{\sqrt{A_{-}^2- 1} }
\end{pmatrix}
\right].
\end{aligned}
\label{gamma_values_749}
\end{equation}
Here
\begin{gather}
\begin{aligned}
A_{\pm} =
\frac
{
	t\mu
	\pm
	\left|\Delta\right| \sqrt{ \mu^2 + 4 (\left|\Delta\right|^2 - t^2) \left(1 - \frac{(\omega + i\delta)^2}{4 \left|\Delta\right|^2} \right) }
}{
	2 (\left|\Delta\right|^2 - t^2)
},
\\
\delta \rightarrow +0,
\end{aligned}
\\
\chi_{\pm} = A_{\pm} - \sqrt{A_{\pm}^2 - 1}.
\end{gather}
The complex value square root $\sqrt{A_{\pm}^2 - 1}$ is defined in a way that it has a branch cut at $A_{\pm}\in (-1, 1)$ and has positive values for $A_{\pm} > 1$.

\begin{figure}
\includegraphics[width=0.4\textwidth]{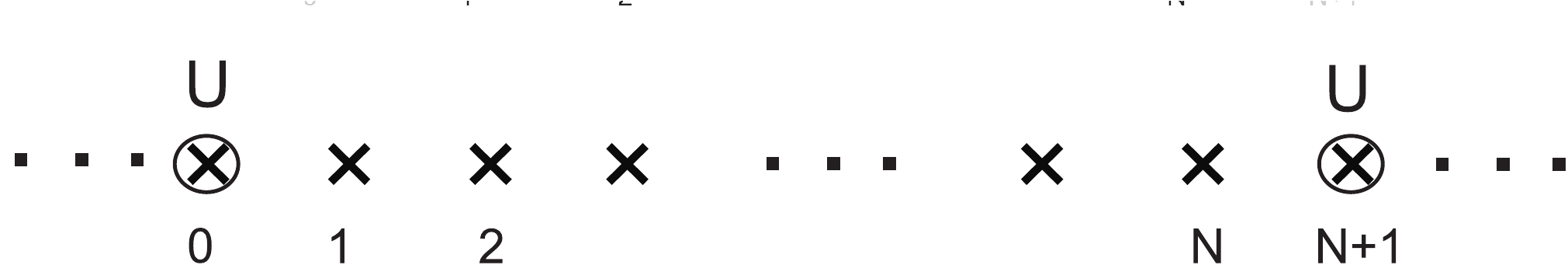}
\caption{Infinite Kitaev chain with two defects}
\end{figure}

The Green's function of Hamiltonian (\ref{ham_orig_short}) can be written in terms of the infinite chain's Green's function using the Dyson equation with perturbation $\hat V$,
\begin{equation}
\begin{aligned}
&\Gamma^{R}_{nm} (\omega) =
	\Gamma^{0R}_{nm} (\omega)
	+
	\Gamma^{0R}_{n0} (\omega) U \sigma_{z} \Gamma^{R}_{0m} (\omega)
\\&
	+
	\Gamma^{0R}_{n,N+1} (\omega) U \sigma_{z} \Gamma^{R}_{N+1,m} (\omega).
\end{aligned}
\label{gamma_orig_932}
\end{equation}
Solving equation (\ref{gamma_orig_932}) for $\Gamma^{R}_{nm} (\omega)$ and accounting for the limit $U \rightarrow \infty$, we find the exact solution for Green's functions $\Gamma^{R}_{nm} (\omega)$:
\begin{widetext}
\begin{equation}
\begin{aligned}
\Gamma^{R}_{nm} (\omega) =&~
	\Gamma^{0R}_{nm} (\omega)
	-
	\Gamma^{0R}_{n0} (\omega)
	\left(
		 \Gamma^{0R}_{0,0} (\omega)
		- \Gamma^{0R}_{0,N+1} (\omega) ( \Gamma^{0R}_{N+1,N+1} (\omega) )^{-1} \Gamma^{0R}_{N+1,0} (\omega)
	\right)^{-1}
\\&\times
	\left(\Gamma^{0R}_{0,m} (\omega) - \Gamma^{0R}_{0,N+1} (\omega) ( \Gamma^{0R}_{N+1,N+1} (\omega) )^{-1} \Gamma^{0R}_{N+1,m} (\omega) \right)
\\&-
	\Gamma^{0R}_{n,N+1} (\omega)
	\left(
		\Gamma^{0R}_{N+1,N+1} (\omega)
		- \Gamma^{0R}_{N+1,0} (\omega) ( \Gamma^{0R}_{00} (\omega) )^{-1} \Gamma^{0R}_{0,N+1} (\omega)
	\right)^{-1}
\\&\times
	\left( \Gamma^{0R}_{N+1,m} (\omega) - \Gamma^{0R}_{N+1,0} (\omega) ( \Gamma^{0R}_{00} (\omega) )^{-1} \Gamma^{0R}_{0m} (\omega) \right).
\end{aligned}
\label{gamma_164}
\end{equation}
\end{widetext}
The elements of matrix $\Gamma^{R}_{nm} (\omega)$ describe Green's functions of the finite chain if the arguments satisfy relation \nobreak{$0 < n, m < N+1$.} It can be checked directly that $\Gamma^{R}_{nm} (\omega) = 0$ if one of the arguments $n$ and $m$ is positive and the other is negative, which gives us the direct proof that our procedure effectively removes site $n = 0$ from the system. The same is true for site $n=N+1$.

One could see that function $\Gamma^{R}_{nm} (\omega)$ may have a set of poles at the values of $\omega$ given by equation
\begin{equation}
	\det \left(
		\Gamma^{0R}_{00} (\omega)
		- \Gamma^{0R}_{N+1,0} (\omega) ( \Gamma^{0R}_{00} (\omega) )^{-1} \Gamma^{0R}_{0,N+1} (\omega)
	\right)
= 0.
\label{det_gamma_164}
\end{equation}
Since $\Gamma^{0R}_{nm} (\omega)$ does not have the poles inside the gap, we can assume that the solutions of this equation correspond to the energies of the states localized at the edges of the chain. The direct substitution of Green's functions \ref{gamma_values_749} allows us to find the solution for $\omega$ for arbitrary values of the parameters.

For a semi-infinite chain, if $N\rightarrow \infty$, the situation is much simplified.
 The equation (\ref{det_gamma_164}) has only a single solution in the gap at
\begin{equation}
	\omega = 0.
\end{equation}
 This solution satisfies (\ref{det_gamma_164}), and thus corresponds to a pole of the Green's functions if
\begin{equation}
\left|\mu\right| < 2t.
\end{equation}
Let us stress, that the solution $\omega = 0$ exists for any set of parameters under this condition. In particular, this solution does not require $\mu$ to be exactly zero as it is usually assumed.
The details of the calculations and some discussion of the case of a semi-infinite chain are given in Appendix \ref{appen}.

If we now consider a long finite chain of length $N$, then we can write the equation for the localized states as
\begin{equation}
\det \left[ \Gamma^{(X)R}_{N+1, N+1} (\omega)\right]
= 0,
\label{eq_full_737}
\end{equation}

where $ \Gamma^{(X)R}_{N+1, N+1} (\omega)$ is the Green's function of the semi-infinite chain, which has been defined in (\ref{gamma_half_217}).
If $N\rightarrow \infty$, then, as follows from the properties of the semi-infinite chain, there are two bound states localized at the edges of the chain with the energy $\omega \rightarrow 0$. Expanding (\ref{eq_full_737}) with respect to $\omega$ and $\chi_{\pm}^{N}$, which we consider to be exponentially small, we get
\begin{widetext}
\begin{equation}
  \label{omega_chain}
\det \left[
	-
	\omega
	\frac{t}{\left|\Delta\right| (4 t^2 - \mu^2)}
	\hat{\text I}
	-
	\frac{1}{\omega}
	\frac
	{
		\left|\Delta\right| (4 t^2 - \mu^2)
	}{
		2t ( 4 ( t^2 - \left|\Delta\right|^2 ) - \mu^2 )
	}
	\left(\chi_{+}^{N+1} - \chi_{-}^{N+1}\right)^2
	\begin{pmatrix}
		1
	&
		\frac{\Delta}{\left|\Delta\right|}
	\\
		\frac{\Delta^{*}}{\left|\Delta\right|}
	&
		1
	\end{pmatrix}
\right] = 0,
\end{equation}
\end{widetext}
where $\hat{\text I}$ is the identity matrix.
The solution $\omega=0$ belongs to the pole of the Green's function that only exists at the semi-infinite parts of the chain.
Another pair of solutions have finite but exponentially small energies
$
\omega = \pm \omega_0,
$
where
\begin{equation}
   \label{omega_0}
\omega_0 =
	\frac
		{\left|\Delta\right| (4 t^2 - \mu^2)}
		{i t \sqrt{ 4 ( t^2 - \left|\Delta\right|^2 ) - \mu^2 } }
	\left(\chi_{+}^{N+1} - \chi_{-}^{N+1}\right).
\end{equation}
Here we can see that this solution satisfies the approximations we have made if $\left|\chi_{\pm}^{N+1}\right|\ll 1 $.
For the two cases $\left|\Delta\right|<<t$ and $\left|\Delta\right|<t, \left|\Delta\right|\to t$, equation (\ref{omega_0}) simplifies to
\begin{equation}
          \label{Omega_0}
\omega_0 =
	\left\{
		\begin{aligned}
		& 4 \left|\Delta\right| e^{-N(\left|\Delta\right|/t)},&&  \left|\Delta\right|<<t,
				\\
			& 2t e^{-N \ln(\sqrt{2t/(t-\left|\Delta\right|)})},&&  (t-\left|\Delta\right|)\ll t.
				\end{aligned} \right.
\end{equation}
As the consequence of the relation $\left|\chi_{\pm}\right|< 1 $, energy $\omega_0$ of the gap state always exponentially decreases with the increase of the chain length $N$. In what follows we will call the terms containing $\left|\chi_{\pm}\right|^N $  "exponentially" small compared to $\left|\chi_{\pm}\right|$.

The leading term of the expansion of the Green's function $\Gamma^{R}_{nm} (\omega)$ at $\omega\rightarrow\pm \omega_{0}$, which in quantum mechanics would correspond to the structure of two localized states, takes the following form:
\begin{widetext}
\begin{equation}
\begin{aligned}
\Gamma^{R}_{nm} (\omega) =&
	-
	\frac
	{
		\omega
	}{
		(\omega + i\delta)^2
		-
		(\omega_{0})^2
	}
	\cdot
	\frac
		{\left|\Delta\right| (4 t^2 - \mu^2) }
		{2 t ( 4 (t^2 - \left|\Delta\right|^2) - \mu^2 )}
\\ &\times
	\left(
		(\chi_{+}^{n} - \chi_{-}^{n})
		(\chi_{+}^{m} - \chi_{-}^{m})
		\begin{pmatrix}
			1
		&
			\frac{\Delta}{\left|\Delta\right|}
		\\
			\frac{\Delta^{*}}{\left|\Delta\right|}
		&
			1
		\end{pmatrix}
	+
		(\chi_{+}^{N+1-n} - \chi_{-}^{N+1-n})
		(\chi_{+}^{N+1-m} - \chi_{-}^{N+1-m})
		\begin{pmatrix}
			1
		&
			- \frac{\Delta}{\left|\Delta\right|}
		\\
			- \frac{\Delta^{*}}{\left|\Delta\right|}
		&
			1
		\end{pmatrix}
	\right)
\\ &-
	\frac
	{
		\omega_{0}
	}{
		(\omega + i\delta)^2
		-
		(\omega_{0})^2
	}
	\cdot
	\frac
		{\left|\Delta\right| (4 t^2 - \mu^2) }
		{2 t ( 4 (t^2 - \left|\Delta\right|^2) - \mu^2 )}
\\ &\times
	\left(
		(\chi_{+}^{n} - \chi_{-}^{n})
		(\chi_{+}^{N+1-m} - \chi_{-}^{N+1-m})
		\begin{pmatrix}
			1
		&
			- \frac{\Delta}{\left|\Delta\right|}
		\\
			\frac{\Delta^{*}}{\left|\Delta\right|}
		&
			- 1
		\end{pmatrix}
	+
		(\chi_{+}^{N+1-n} - \chi_{-}^{N+1-n})
		(\chi_{+}^{m} - \chi_{-}^{m})
		\begin{pmatrix}
			1
		&
			\frac{\Delta}{\left|\Delta\right|}
		\\
			- \frac{\Delta^{*}}{\left|\Delta\right|}
		&
			- 1
		\end{pmatrix}
	\right).
\end{aligned}
\label{state_454}
\end{equation}
The Green's function in this case turns into a relatively simple form:
\begin{equation}
   \label{GammaR0}
\begin{aligned}
\Gamma^{R}_{nm} (\omega \rightarrow \pm \omega_0) =&
	-
	\frac
	{
		1
	}{
		\omega
		\mp
		\omega_{0}
		+ i\delta
	}
	\cdot
	\frac
		{\left|\Delta\right| t }
		{ t^2 - \left|\Delta\right|^2 }
\\ &\times
	\left[
		\left(
			\frac
			{
				t - \left|\Delta\right|
			}{
				t + \left|\Delta\right|
			}
		\right)^{\frac{n+m}{2}}
		\sin \left( \frac{\pi n}{2} \right)
		\sin \left( \frac{\pi m}{2} \right)
		\begin{pmatrix}
			1
		&
			\frac{\Delta}{\left|\Delta\right|}
		\\
			\frac{\Delta^{*}}{\left|\Delta\right|}
		&
			1
		\end{pmatrix}
\right.\\ &~~~~+\left.
		\left(
			\frac
			{
				t - \left|\Delta\right|
			}{
				t + \left|\Delta\right|
			}
		\right)^{N+1-\frac{n+m}{2}}
		\sin \left( \frac{\pi (N+1-n)}{2} \right)
		\sin \left( \frac{\pi (N+1-m)}{2} \right)
		\begin{pmatrix}
			1
		&
			- \frac{\Delta}{\left|\Delta\right|}
		\\
			- \frac{\Delta^{*}}{\left|\Delta\right|}
		&
			1
		\end{pmatrix}
\right.\\ &~~~~\pm \left.
		\left(
			\frac
			{
				t - \left|\Delta\right|
			}{
				t + \left|\Delta\right|
			}
		\right)^{\frac{N+1+n-m}{2}}
		\sin \left( \frac{\pi n}{2} \right)
		\sin \left( \frac{\pi (N+1-m)}{2} \right)
		\begin{pmatrix}
			1
		&
			- \frac{\Delta}{\left|\Delta\right|}
		\\
			\frac{\Delta^{*}}{\left|\Delta\right|}
		&
			- 1
		\end{pmatrix}
\right.\\ &~~~~\pm \left.
		\left(
			\frac
			{
				t - \left|\Delta\right|
			}{
				t + \left|\Delta\right|
			}
		\right)^{\frac{N+1-n+m}{2}}
		\sin \left( \frac{\pi (N+1-n)}{2} \right)
		\sin \left( \frac{\pi m}{2} \right)
		\begin{pmatrix}
			1
		&
			\frac{\Delta}{\left|\Delta\right|}
		\\
			- \frac{\Delta^{*}}{\left|\Delta\right|}
		&
			- 1
		\end{pmatrix}
	\right].
\end{aligned}
\end{equation}
\end{widetext}
The diagonal elements $\Gamma_{nn}$ show the spatial distribution of the localized states. In the limit $\Delta=t$ only $\Gamma_{11}$ and $\Gamma_{NN}$ are not equal to zero.

 In the highly symmetrical case $\mu = 0$ and $\left|\Delta\right| \rightarrow t$ the the energy levels is equal to:
\begin{equation}
\omega_0 =
	\frac
		{4 \left|\Delta\right| t}
		{ t + \left|\Delta\right| }
	\left(
		\frac
		{
			t - \left|\Delta\right|
		}{
			t + \left|\Delta\right|
		}
	\right)^{\frac{N}{2}}
	\sin \left( \frac{\pi (N+1)}{2} \right)
	\rightarrow 0.
\end{equation}
As it was noticed before (e.g.\cite{even_odd}), for the odd number of sites in the chain $\omega_0$ is equal to zero for arbitrary values of $t$ and $\Delta$ .

In the limit $N\rightarrow \infty$, the states around each edge of the chain start to behave as if the chain is semi-infinite. Naively these two poles with residues equal to $1/2$ correspond to a "particle" which is split between two states. However, the very fact that the energy of the ''localized state'' is always equal to zero and does not depend on the parameters $\mu, t, \Delta$ (if $|\mu| < 2t$) means that it is not a real single-particle localized state (excitation). As it was mentioned in the paper \cite{kit} the ground state of the model is doubly degenerate. The two ground states differ by the number of fermions, so the matrix element of creation or annihilation operator between these ground states is not zero, which gives us the pole in the Green function at zero energy. Note that all degenerate ground states give equally weighted contributions to all average values. That explains the appearance of $1/2$ residue at the end of a semi-infinite chain.

This statement can be easily illustrated in a simple example of the two-site chain.
The Hamiltonian (\ref{ham_zero_000}) for two sites can be easily diagonalized by Bogolubov transformation. In terms of Bogolubov operators the Hamiltonian takes the form
\begin{equation}
\hat H_{2} =E_0+\varepsilon_1 c_1^{\dagger}c_1+\varepsilon_2 c_2^{\dagger}c_2,
\label{ham_2}
\end{equation}
where
\begin{gather*}
 c_{1,2}=u \frac{( \psi_1+ \psi_2)}{\sqrt{2}}\pm v \frac{(\psi^{\dagger}_{1}-\psi^{\dagger}_{2})}{\sqrt{2}},
\\
v^2,u^2=\frac{1}{2}\left(1\pm\frac{\mu}{t}\right),
\\
\varepsilon_{1,2}= t\left(1 \mp \frac{\Delta}{\sqrt{t^2-\mu^2}}\right).
\end{gather*}
For ${\Delta}={\sqrt{t^2-\mu^2}} $ (if  $ \mu=0$  it gives $\Delta=t$)
 one gets $\varepsilon_1=0$ (this is the solution of (\ref{omega_chain}) for $N=2$ and $\omega=0$).
Then
 \begin{equation*}
\left|\Phi_0\right>=\frac{1}{\sqrt 2}(\psi^{\dagger}_{1}-\psi^{\dagger}_{2})\left|0\right>.
\end{equation*}
is the ground state which satisfies
$$
c_{1,2}\left|\Phi_0\right>=0
$$
At the same time, the state
\begin{equation*}
\left|\Phi_1\right>= c_1^{\dagger} \left|\Phi_0\right>  =(v+u\psi^{\dagger}_{1}\psi^{\dagger}_{2} ) \left|0\right>
\end{equation*}
also has zero energy, which means that the ground state is degenerate.
As to particle matrix elements between these two ground states, we have:
\begin{align*}
\left<\Phi_0\right|\psi_1\left|\Phi_1\right>= u/\sqrt{2},
&&
  \left<\Phi_1\right|\psi_1\left|\Phi_0\right>= v/\sqrt{2}.
\end{align*}
This means that in the single-particle function $G_{11}$, a pole with residue equal to $1/2$ appears at $\omega=0$.

\section{Tunneling current}

Let us now study the stationary tunneling properties of a Kitaev chain. The physically reasonable way to do this is to assume that the chain is connected at the sites $1$ and $N$ to two external reservoirs (contact leads) with a large number of degrees of freedom, denoted with indices $l$ and $r$, respectively. The complete Hamiltonian then can be written as
\begin{equation}
     \label{Htun}
\begin{aligned}
\hat{\tilde H} =&~
	\hat H
	+
	\sum_{p} \tau^{l}_{p}
	\left(
		h^{l\dagger}_{p} \psi_{1} + \psi^{\dagger}_{1} h^{l}_{p}
	\right)
\\&+
	\sum_{p} \tau^{r}_{p}
	\left(
		h^{r\dagger}_{p} \psi_{N} + \psi^{\dagger}_{N} h^{r}_{p}
	\right)
\\&+
	\sum_{p}
	E^{l}_{p} h^{l\dagger}_{p} h^{l}_{p}
	+
	\sum_{p}
	E^{r}_{p} h^{r\dagger}_{p} h^{r}_{p}.
\end{aligned}
\end{equation}
The current flowing into the chain through site $1$ is determined in the usual way as (\cite{CarCombNoz})
\begin{equation}
    \label{Itun1}
I_{l} (t) =
i \sum_{p} \tau^{l}_{p} \left<
	h^{l\dagger}_{p} \psi_{1} - \psi^{\dagger}_{1} h^{l}_{p}
\right>.
\end{equation}
Using the non-stationary diagram technique formalism, this expression can be rewritten as
\begin{equation}
I_{l} (t) =
- \sum_{p} \tau^{l}_{p}
\left(
	\tilde G^{<}_{lp, 1} (t, t) - \tilde G^{<}_{1, lp} (t, t)
\right),
\label{cur_start_t_7529}
\end{equation}
where
\begin{equation*}
\begin{aligned}
\tilde G^{<}_{lp, 1} (t, t) =&
	\int dt_1 g^{<}_{lp} (t, t_1) \tau^{l}_{p} \tilde G^{A}_{1, 1} (t_1, t)
\\&+
	\int dt_1 g^{R}_{lp} (t, t_1) \tau^{l}_{p} \tilde G^{<}_{1, 1} (t_1, t),
\\
\tilde G^{<}_{1, lp} (t, t) =&
	\int dt_1 \tilde G^{<}_{1, 1} (t, t_1) \tau^{l}_{p} g^{A}_{lp} (t_1, t)
\\&+
	\int dt_1 \tilde G^{R}_{1, 1} (t, t_1) \tau^{l}_{p} g^{<}_{lp} (t_1, t).
\end{aligned}
\end{equation*}
Parameter $p$ numerates the modes inside each reservoir; $g_{\alpha p} (\omega)$ is a Green's function in reservoir $\alpha$, where $\alpha$ is $l$ or $r$, when it is not connected to the chain; $\tilde G^{R}_{n,m} (t, t_1)$ are the exact Green's functions of the chain with the external reservoirs taken into account.

It is very important, that the tunneling Hamiltonian (\ref{Htun}) and the tunneling current (\ref{Itun1}) are expressed in terms of real electron operators so you always get the correct electrical current in the answer. Attempts to use from the very beginning  some effective equations in terms of quasipartical (or even "Majorana") operators  often give absolutely incorrect results (as e.g. in Ref.\cite{smirnov}).

We assume as usual that, due to the large number of particles and degrees of freedom in each reservoir, the particle distribution function does not change substantially throughout the whole experiment and consequently that each reservoir remains almost in the equilibrium state. The system as a whole, however, is not in equilibrium, though in this section we consider it to be stationary, and the value of the current does not change with time. Thus, expression (\ref{cur_start_t_7529}) can be rewritten using the frequency-dependent Green's functions as follows:
\begin{equation}
I_{l} =
- \sum_{p} \tau^{l}_{p}
\int \frac{d\omega}{2\pi}
\left(
	\tilde G^{<}_{lp, 1} (\omega) - \tilde G^{<}_{1, lp} (\omega)
\right).
\label{current_omega_385}
\end{equation}
Here
\begin{equation}
\begin{aligned}
\tilde G^{<}_{lp, 1} (\omega) &=
	g^{<}_{lp} (\omega) \tau^{l}_{p} \tilde G^{A}_{1, 1} (\omega)
	+
	g^{R}_{lp} (\omega) \tau^{l}_{p} \tilde G^{<}_{1, 1} (\omega),
\\
\tilde G^{<}_{1, lp} (t, t) &=
	\tilde G^{<}_{1, 1} (\omega) \tau^{l}_{p} g^{A}_{lp} (\omega)
	+
	\tilde G^{R}_{1, 1} (\omega) \tau^{l}_{p} g^{<}_{lp} (\omega).
\end{aligned}
\end{equation}

We can simplify this expression if we define self-energy
\begin{equation}
\Sigma^{R (A, <)}_{\alpha} (\omega) =
	\sum_{p} \left(\tau^{\alpha}_{p}\right)^2 g^{R (A, <)}_{\alpha p} (\omega).
\end{equation}
Then we can take into account the fact that each reservoir is kept in equilibrium by equation
\begin{equation}
\Sigma^{<}_{\alpha} (\omega) = n_{\alpha} (\omega) \left(\Sigma^{A}_{\alpha} (\omega) - \Sigma^{R}_{\alpha} (\omega) \right),
\end{equation}
where $n_{\alpha} (\omega)$ is the Fermi distribution function for reservoirs $l$ and $r$.
Then equation (\ref{current_omega_385}) can be rewritten as
\begin{equation}
     \label{I1}
\begin{aligned}
\hat I_{l} =&
	- \int \frac{d\omega}{2\pi}
	\left(\Sigma^{A}_{l} (\omega) - \Sigma^{R}_{l} (\omega) \right)
\\&\times
	\left(
		n_{l} (\omega) \left(\tilde \Gamma^{A}_{1, 1} (\omega) - \tilde \Gamma^{R}_{1, 1} (\omega) \right)
		-
		\tilde \Gamma^{<}_{1, 1} (\omega)
	\right).
\end{aligned}
\end{equation}
Here current $I_{l}$ is given by the upper-left element of matrix $\hat I_{l}$ ($\hat I^{11}$). Expression of this type in terms of non-equilibrium Green functions was firstly derived in \nobreak{Ref. \cite{CarCombNoz}} and later was applied in Ref. \cite{Meir}. This expression seems to be asymmetric with respect to the left and right leads. But for the stationary case, correctly calculated current  (\ref{I1}) always can be written in an explicitly symmetric form.
In our case, equation (\ref{I1}) can be further simplified using relations
\begin{widetext}
\begin{gather}
\tilde \Gamma^{<}_{1, 1} (\omega) =
	\tilde \Gamma^{R}_{1, 1} (\omega)
	\Sigma^{<}_{l} (\omega)
	\tilde \Gamma^{A}_{1, 1} (\omega)
	+
	\tilde \Gamma^{R}_{1, N} (\omega)
	\Sigma^{<}_{r} (\omega)
	\tilde \Gamma^{A}_{N, 1} (\omega),
\\
\tilde \Gamma^{A}_{1, 1} (\omega) - \tilde \Gamma^{R}_{1, 1} (\omega)=
	\tilde \Gamma^{R}_{1, 1} (\omega)
	\left( \Sigma^{A}_{l} (\omega) - \Sigma^{R}_{l} (\omega) \right)
	\tilde \Gamma^{A}_{1, 1} (\omega)
	+
	\tilde \Gamma^{R}_{1, N} (\omega)
	\left( \Sigma^{A}_{r} (\omega) - \Sigma^{R}_{r} (\omega) \right)
	\tilde \Gamma^{A}_{N, 1} (\omega).
\end{gather}
\end{widetext}
Going further, we use broad band approximation for the reservoirs, which means that for the values of $\omega$ under consideration
\begin{align*}
&\Sigma^{A}_{l (r)} (\omega) \approx i \gamma_{l (r)},
&
&\Sigma^{R}_{l (r)} (\omega) \approx - i \gamma_{l (r)}.
\end{align*}

where $\gamma_{l (r)}=\pi \nu^{l (r)}(\tau^{l (r)}_{p})^2$ and  $\nu^{l (r)} $ is the density of states in reservoir $l (r)$.

The direct substitution yields
\begin{equation}
\hat I_{l} =
	4\gamma_{l}\gamma_{r} \int \frac{d\omega}{2\pi}
	\tilde \Gamma^{R}_{1, N} (\omega)
	\tilde \Gamma^{A}_{N, 1} (\omega)(n_{l} (\omega) - n_{r} (\omega)),
\label{current_greens_final_205}
\end{equation}

A formula of this type was at first derived in Ref. \cite{CarCombNoz}.
We should note that the resulting equation for the current through a system is always symmetric for the system two edges. It means that, in the stationary case, the current flowing into the system is the same as the one flowing out of it. Conservation of the total current can not be violated in any system whatever "topologically usual" it is and does not require some additional conditions like equal tunneling rates to the leads and symmetrically applied bias. So the appearance of non-symmetric expressions for the stationary tunneling current obtained in some papers (e.g.\cite{leumer}) concerning systems like Kitaev chain is simply the result of incorrect calculations. This statement does not change for systems with interaction, but the derivation of explicitly symmetric expression is more difficult in this case. The examples of such calculation for systems with electron-phonon interaction are given e.g. in \cite{CarCombNoz4, UFN10}.
Let us stress that the equation (\ref{current_greens_final_205}) is exact and explicitly symmetric with respect to the left and the right leads.

Since we aim to explore the properties of the low-energy resonance, we need to make sure that only this resonance is responsible for the transfer of particles through the chain. It is possible to achieve this if we assume that $\delta n (\omega)=n_{l} (\omega) - n_{r} (\omega)$ is non-zero only for the values of $\omega$ lying within the superconducting gap.
In other words, the applied voltage should be less than the superconducting gap.
In this case, we eliminate the influence of the states in the continuous spectrum.

To express $\widetilde \Gamma^{R}_{1,N} (\omega)$ in terms of $\Gamma^{R}_{n,m} (\omega)$, we use Dyson equation
\begin{equation}
\begin{aligned}
\widetilde \Gamma^{R}_{n, m} (\omega) =&
	\Gamma^{R}_{n, m} (\omega)
	+
	\Gamma^{R}_{n, 1} (\omega)
	\Sigma^{R}_{l} (\omega)
	\widetilde \Gamma^{R}_{1, m} (\omega)
\\&+
	\Gamma^{R}_{n, N} (\omega)
	\Sigma^{R}_{r} (\omega)
	\widetilde \Gamma^{R}_{N, m} (\omega).
\end{aligned}
\end{equation}
Simple algebraic transformations show that
\begin{widetext}
\begin{equation}
\begin{aligned}
\widetilde \Gamma^{R}_{1, N} (\omega) =&
	\left[
		\hat{\text I}
		+
		\gamma_{l}\gamma_{r}
		\left( \hat{\text I} + i \gamma_{l} \Gamma^{R}_{1, 1} (\omega) \right)
		\Gamma^{R}_{1, N} (\omega)
		\left( \hat{\text I} + i \gamma_{r} \Gamma^{R}_{N, N} (\omega) \right)
		\Gamma^{R}_{N, 1} (\omega)
	\right]^{-1}
\\&\times
	\left( \hat{\text I} + i \gamma_{l} \Gamma^{R}_{1, 1} (\omega) \right)
	\Gamma^{R}_{1, N} (\omega)
	\left( \hat{\text I} + i \gamma_{r} \Gamma^{R}_{N, N} (\omega) \right).
\end{aligned}
\end{equation}
\end{widetext}
where $\hat{\text I}$ is the identity matrix.

In the case $\gamma_{l,r}>\Delta$, we can not eliminate the contribution to the current from continuous states and all information about localized states is lost.
So the following calculations are made for the case with the following hierarchy of parameters: $t>\Delta >\gamma_{l,r} $. Analyzing
different terms in (\ref{state_454}) for $\Gamma_{nm}$, we find that the simplest form can be obtained if parameter $\Delta^2/(t\gamma)\gg 1$.
Retaining the leading terms in this parameter, we get the result

\begin{equation}
\begin{aligned}
   \label{Gamma1N}
\widetilde \Gamma^{R}_{1, N} (\omega) = &~
	\frac
	{
		C \omega_{0}
	}{
		\omega^2 - \omega_0^2 + 2 i ( \gamma_{l} + \gamma_{r} ) C \omega - 4\gamma_{l} \gamma_{r} C^2
	}
\\&\times
	\begin{pmatrix}
		1
	&
		- \frac{\Delta}{\left|\Delta\right|}
	\\
		\frac{\Delta^{*}}{\left|\Delta\right|}
	&
		- 1
	\end{pmatrix},
\end{aligned}
\end{equation}
where
\begin{equation}
         \label{C}
\begin{aligned}
C \equiv &
	- \frac
	{
		\left|\Delta\right| (4t^2 - \mu^2)
	}{
		2t (4 (t^2 - \left|\Delta\right|^2 ) - \mu^2 )
	}
	(\chi_{+} - \chi_{-} )^2
\\=&~
	\frac
	{
		\left|\Delta\right| (4t^2 - \mu^2)
	}{
		2t (t+ \left|\Delta\right|)^2
	}.
\end{aligned}
\end{equation}

Substituting this result into (\ref{current_greens_final_205}), we now get
\begin{widetext}
\begin{equation}
I_{l} =
	\int \frac{d\omega}{2\pi}
	\frac
	{
		8 \gamma_{l}\gamma_{r} C^2 \omega_{0}^2
	}{
		(\omega^2 - \omega_0^2 + 2 i ( \gamma_{l} + \gamma_{r} ) C \omega - 4\gamma_{l} \gamma_{r} C^2)
		(\omega^2 - \omega_0^2 - 2 i ( \gamma_{l} + \gamma_{r} ) C \omega - 4\gamma_{l} \gamma_{r} C^2)
	}(n_{l} (\omega) - n_{r} (\omega)).
\label{cur_comp_93290}
\end{equation}
\end{widetext}
The value of the current is proportional to $\omega_{0}^2$, so it decreases exponentially with the increase of the length of the chain. Moreover, if $\omega_0=0$, which is the case generally associated with Majorana particles, the current through the system does not flow at all. Note that equation (\ref{cur_comp_93290}) is symmetrical for leads $l$ and $r$ as it should be.

In particular, if the applied voltage is greater than the width of the localized states but less than the superconducting gap value, which gives $n_{l} (\omega) - n_{r} (\omega) = 1$ for $\left|\omega\right| \lesssim \gamma_{l}, \gamma_{r}$, then we get the simple final expression for the tunneling current through the chain associated with "Majorana" states:

\begin{equation}
I_{l} =
	\frac{2 \gamma_{l}\gamma_{r} C \omega_{0}^2}{\gamma_{l} + \gamma_{r} }
	~
	\frac
	{
		1
	}{
		4 \gamma_{l} \gamma_{r} C^2 + \omega_{0}^2
	}.
\label{cur_comp_93290}
\end{equation}
So the value of the current is always determined by the smallest rate parameter (the weak link of the system); in our case, these parameters are $\omega_0^2/(\gamma_{l} + \gamma_{r}),\gamma_{l}, \gamma_{r}$.
 If $\omega_0^2\ge C^2\gamma_{l}\gamma_{r}$, then general equation (\ref{cur_comp_93290}) leads to the current value proportional to $\gamma_{l}\gamma_{r}/(\gamma_{l} + \gamma_{r})$, which is the usual coefficient for the tunneling through an intermediate state in a common tunneling problem.
For this system, relation $\omega_0\ll \gamma_{l}, \gamma_{r}$ is physically reasonable. Using it we obtain:
\begin{equation}
I_{l} =
	\frac{ \omega_{0}^2}{2 C (\gamma_{l} + \gamma_{r}) }.
\end{equation}
If we use equations (\ref{Omega_0}) and (\ref{C}), then this formula gives:
\begin{equation}
          \label{Istat}
I =
	\left\{
		\begin{aligned}
		& \frac{ 2\Delta t }{(\gamma_{l} + \gamma_{r}) }e^{-2N(\Delta/t)}
              ,&  \Delta&\ll t,
              \\
	& \frac{ 2 t^2 }{(\gamma_{l}+ \gamma_{r}) } e^{-N \ln(2t/(t-\Delta))}
                ,&  (t-\Delta)&\ll t.
				\end{aligned} \right.
\end{equation}
For arbitrary parameters $\mu <\Delta <t$, the current is always exponentially small for long chains. Let us repeat, in the case $\omega_0=0$, which is considered the most favorable case for unusual topological properties, we can not see any tunneling conductivity peak at zero voltage at all.

Naively used common formulas for the tunneling current between two leads are misleading for low dimensional systems and objects like Kitaev chain \cite{zazunov} because for these objects we always have to define the physical procedure for fixing the chemical potential. We can not do it otherwise than make a contact between some reservoir and the system. If we make such additional contact to the Kitaev chain we immediately create the edge state  and  this problem of two contacts just has been solved here.

\section{Non-stationary current}

We now try to answer the question what the typical time scales for the current or charge transfer from one edge of the chain to the other one are.

To do this we will assume that the system as a whole is at equilibrium at $t<0$, and then at $t=0$ bias is applied to one of the leads. This additional bias causes a non-stationary current, which at $t\rightarrow + \infty$ reaches the stationary value (\ref{cur_comp_93290}).

Applied bias increases the energy levels in the reservoirs by $V_{\alpha}$, where index $\alpha$ designates the reservoir. Thus, the Hamiltonian of the reservoirs can now be written as
\begin{equation}
\begin{aligned}
\hat H_{\alpha} (t) = &
	\sum_{p} \tau^{\alpha}_{p}
	\left(
		h^{\alpha\dagger}_{p} \psi_{1} + \psi^{\dagger}_{1} h^{\alpha}_{p}
	\right)
\\&+
	\sum_{p}
	\left(E^{\alpha}_{p} + V_{\alpha} \theta (t) \right) h^{\alpha\dagger}_{p} h^{\alpha}_{p}.
\end{aligned}
\end{equation}
For now, we focus on the "left" reservoir {\it l} (for the "right" lead {\it r} all the formulas can be written the same way). The current flowing from the left reservoir into the system is given by
\begin{equation}
\begin{aligned}
I_{l} (t) =
- \int dt'
&\left(
	\Sigma^{<}_{l} (t, t') \widetilde G^{A}_{1, 1} (t', t)
	+ \Sigma^{R}_{l} (t, t_1) \widetilde G^{<}_{1, 1} (t', t)
\right.\\&~ \left.
	- \widetilde G^{<}_{1, 1} (t, t') \Sigma^{A}_{l} (t', t)
	- \widetilde G^{R}_{1, 1} (t, t') \Sigma^{<}_{l} (t', t)
\right).
\end{aligned}
\label{nonstatcur121}
\end{equation}
Here the self-energy part is defined as
\begin{gather}
\begin{aligned}
\Sigma^{R}_{\alpha} (t, t') = &
	-i \sum_{p} (\tau^{\alpha}_{p})^2  \theta (t - t')
\\&\times
	\exp \left( -i E^{\alpha}_{p} (t - t') - i V_{\alpha} \int_{t'}^{t} dt_1 \theta (t_1) \right),
\end{aligned}
\\
\begin{aligned}
\Sigma^{<}_{\alpha} (t, t') = &~
	i \sum_{p} (\tau^{\alpha}_{p})^2 n^{\alpha}_{p}
\\&\times
	\exp \left( -i E^{\alpha}_{p} (t - t') - i V_{\alpha} \int_{t'}^{t} dt_1 \theta (t_1) \right).
\end{aligned}
\end{gather}
In the frequency representation, these expressions correspond to formulas
\begin{widetext}
\begin{gather}
\begin{aligned}
\Sigma^{R}_{\alpha} (\omega, \omega') =
	-i (\tau^{\alpha})^2 \int d\varepsilon \nu^{\alpha} (\varepsilon)
	&\left[
		- \frac{1}{\omega - \varepsilon - V_{\alpha} + 2i\delta}
		\left(
			 - \frac{1}{\omega' - \omega - 2i\delta}
			 + \frac{1}{\omega' - \omega + V_{\alpha}}
		\right)
\right.\\&~+\left.
		\frac{1}{\omega' - \varepsilon + 2i\delta}
		\left(
			 - \frac{1}{\omega - \omega' - V_{\alpha}}
			 + \frac{1}{\omega - \omega' - 2i\delta}
		\right)
	\right],
\end{aligned}
\\
\label{Sigma<}
\begin{aligned}
\Sigma^{<}_{\alpha} (\omega, \omega') =
	i (\tau^{\alpha})^2 \int & d\varepsilon \nu^{\alpha} (\varepsilon) n^{\alpha} (\varepsilon)
	\left(
		 \frac{1}{\omega - \varepsilon - V_{\alpha} + i\delta}
		 - \frac{1}{\omega - \varepsilon - i\delta}
	\right)
\\&\times
	\left(
		 \frac{1}{\omega' - \varepsilon - V_{\alpha} - i\delta}
		 - \frac{1}{\omega' - \varepsilon + i\delta}
	\right),
\end{aligned}
\end{gather}
\end{widetext}
where $\nu^{\alpha} (\varepsilon)$ is the density of states in reservoir $\alpha$, $\delta\rightarrow +0$. For simplicity, we assume that $\tau^{\alpha}$ does not depend on $p$.
For the broad band approximation, if we assume that $\nu (\varepsilon)$ does not change substantially for $\varepsilon \sim \omega, \omega', V_{l, r}$, these expressions symplify to
\begin{equation*}
\Sigma^{R}_{l} (\omega, \omega') =
	- i \gamma_{l}
	\cdot 2\pi \delta (\omega' - \omega),
\end{equation*}
and
\begin{equation*}
\begin{aligned}
\Sigma^{<}_{l} (\omega, \omega') = &
	\frac{i \gamma_{l}}{\pi} \int d\varepsilon n^{l} (\varepsilon)
	\left(
		 \frac{1}{\omega - \varepsilon - V_{l} + i\delta}
		 - \frac{1}{\omega - \varepsilon - i\delta}
	\right)
\\&\times
	\left(
		 \frac{1}{\omega' - \varepsilon - V_{l} - i\delta}
		 - \frac{1}{\omega' - \varepsilon + i\delta}
	\right).
\end{aligned}
\end{equation*}

As a result, in the frequency representation (\ref{nonstatcur121}) simplifies to
\begin{equation}
\begin{aligned}
\hat I_{l} (\omega) =
- \int \frac{d\Omega}{2\pi} &
	\left(
		\Sigma^{<}_{l} (\Omega, \Omega - \omega)
		\left(\widetilde \Gamma^{A}_{1, 1} (\Omega - \omega) - \widetilde \Gamma^{R}_{1, 1} (\Omega) \right)
\right.\\&~\left.
		- 2i\gamma_{l} \widetilde \Gamma^{<}_{1, 1} (\Omega, \Omega - \omega)
	\right).
\end{aligned}
\label{current_trough_gamma_823}
\end{equation}
Here
\begin{equation}
\begin{aligned}
\widetilde \Gamma^{<}_{1, 1} (\Omega, \Omega - \omega)
	= &\widetilde \Gamma^{R}_{1, 1} (\Omega)  \Sigma^{<}_{l} (\Omega, \Omega - \omega) \widetilde \Gamma^{A}_{1, 1} (\Omega - \omega)
\\&
	+ \widetilde \Gamma^{R}_{1, N} (\Omega)  \Sigma^{<}_{r} (\Omega, \Omega - \omega) \widetilde \Gamma^{A}_{N, 1} (\Omega - \omega).
\end{aligned}
\end{equation}
Using the Dyson equations for retarded and advanced Green's functions, we can also prove that
\begin{widetext}
\begin{equation*}
\widetilde \Gamma^{A}_{1, 1} (\Omega - \omega) - \widetilde \Gamma^{R}_{1, 1} (\Omega) =
	\omega \sum_{n = 1}^{N} \Gamma^{R}_{1, n} (\Omega) \Gamma^{A}_{n, 1} (\Omega - \omega)
	+ \widetilde \Gamma^{R}_{1, 1} (\Omega) \cdot 2i\gamma_{l} \widetilde \Gamma^{A}_{1, 1} (\Omega - \omega)
	+ \widetilde \Gamma^{R}_{1, N} (\Omega) \cdot 2i\gamma_{r} \widetilde \Gamma^{A}_{N, 1} (\Omega - \omega).
\end{equation*}
Substituting these two expressions into (\ref{current_trough_gamma_823}), we get
\begin{equation}
\begin{aligned}
\hat I_{l} (\omega) =
- \int \frac{d\Omega}{2\pi}
&\left(
	\omega \sum_{n = 1}^{N} \widetilde \Gamma^{R}_{1, n} (\Omega) \widetilde \Gamma^{A}_{n, 1} (\Omega - \omega) \Sigma^{<}_{l} (\Omega, \Omega - \omega)
\right.\\&~+\left.
	2i \left(
		\gamma_{r} \Sigma^{<}_{l} (\Omega, \Omega - \omega)
		 - \gamma_{l} \Sigma^{<}_{r} (\Omega, \Omega - \omega)
	\right)
	\widetilde \Gamma^{R}_{1, N} (\Omega) \widetilde \Gamma^{A}_{N, 1} (\Omega - \omega)
\right).
\label{cur_nonstat_green_f_8349}
\end{aligned}
\end{equation}
\end{widetext}
Here
\begin{equation}
\begin{aligned}
&\Sigma^{<}_{r} (\omega, \omega') =
	\frac{i \gamma_{r}}{\pi} \int d\varepsilon n^{l} (\varepsilon)
	\left(
		 \frac{1}{\omega - \varepsilon - V_{r} + i\delta}
		 - \frac{1}{\omega - \varepsilon - i\delta}
	\right)
\\&\times
	\left(
		 \frac{1}{\omega' - \varepsilon - V_{r} - i\delta}
		 - \frac{1}{\omega' - \varepsilon + i\delta}
	\right).
\end{aligned}
\end{equation}

We can see that the first term in (\ref{cur_nonstat_green_f_8349})  exists only if $V_{l}\ne 0$ and does not directly depend on the properties of reservoir {\it r}. This means that this term corresponds to the refilling of the states at the left edge of the chain due to a change in its chemical potential. Consequently, the second term represents the current that flows from one reservoir into the other through the chain. If we now take only the second term into account, then we get
\begin{equation}
\begin{aligned}
\hat I_{l} (t) = &~
	\frac{2 \gamma_{l} \gamma_{r}}{\pi}
	\int d\varepsilon dV
	\hat M_{1, N} (t, \varepsilon, V) \left(\hat M_{1, N} (t, \varepsilon, V) \right)^{\dagger}
\\&\times
	\left[n^{l} (\varepsilon) \delta (V - V_{l}) - n^{r} (\varepsilon) \delta (V - V_{r})\right],
\end{aligned}
\end{equation}
where
\begin{equation*}
\begin{aligned}
\hat M_{1, N} (t, \varepsilon, V) = &
	\int \frac{d\Omega}{2\pi} e^{-i \Omega t}
	\widetilde \Gamma^{R}_{1, N} (\Omega)
\\&\times
	\left(
		 \frac{1}{\Omega - \varepsilon - V + i\delta}
		 - \frac{1}{\Omega - \varepsilon - i\delta}
	\right).
\end{aligned}
\end{equation*}
It can also be shown by direct calculations that
\begin{widetext}
\begin{equation}
\begin{aligned}
\hat M_{1, N} (t, \varepsilon, V) =
	-& i \theta (-t)
	\frac
	{
		e^{-i \varepsilon t}
		C \omega_{0}
	}{
		(\varepsilon + i C (\gamma_{l} + \gamma_{r}))^2 - \overline{\omega}^2
	}
	\begin{pmatrix}
		1
	&
		- \frac{\Delta}{\left|\Delta\right|}
	\\
		\frac{\Delta^{*}}{\left|\Delta\right|}
	&
		- 1
	\end{pmatrix}
	-
	i \theta (t)
	\frac
	{
		e^{-i (\varepsilon + V_{r}) t}
		C \omega_{0}
	}{
		(\varepsilon + V + i C (\gamma_{l} + \gamma_{r}))^2 - \overline{\omega}^2
	}
	\begin{pmatrix}
		1
	&
		- \frac{\Delta}{\left|\Delta\right|}
	\\
		\frac{\Delta^{*}}{\left|\Delta\right|}
	&
		- 1
	\end{pmatrix}
\\-&
	\frac
	{
		 i C \omega_{0} \theta (t)
	}{
		2 \overline{\omega}
	}
	e^{- C ( \gamma_{l} + \gamma_{r}) t - i \overline{\omega} t}
	\begin{pmatrix}
		1
	&
		- \frac{\Delta}{\left|\Delta\right|}
	\\
		\frac{\Delta^{*}}{\left|\Delta\right|}
	&
		- 1
	\end{pmatrix}
	\left(
		- \frac{1}{ \varepsilon + V + i C ( \gamma_{l} + \gamma_{r}) - \overline{\omega} }
		+ \frac{1}{ \varepsilon + i C ( \gamma_{l} + \gamma_{r}) - \overline{\omega} }
	\right)
\\-&
	\frac
	{
		 i C \omega_{0} \theta (t)
	}{
		- 2 \overline{\omega}
	}
	e^{- C ( \gamma_{l} + \gamma_{r}) t + i \overline{\omega} t}
	\begin{pmatrix}
		1
	&
		- \frac{\Delta}{\left|\Delta\right|}
	\\
		\frac{\Delta^{*}}{\left|\Delta\right|}
	&
		- 1
	\end{pmatrix}
	\left(
		- \frac{1}{ \varepsilon + V + i C ( \gamma_{l} + \gamma_{r}) + \overline{\omega} }
		+ \frac{1}{ \varepsilon + i C ( \gamma_{l} + \gamma_{r}) + \overline{\omega} }
	\right).
\end{aligned}
\end{equation}
\end{widetext}
Here
\begin{equation}
    \label{line_omega}
\overline{\omega} = \sqrt{\omega_{0}^2 - C^2 (\gamma_{l} - \gamma_{r})^2}.
\end{equation}
Since we aim to explore how the perturbation transfers through the chain, let us change the bias only at the right lead at the initial time $t=0$ and look at the time-dependent current into the left lead retaining $V_{l} = 0$. Then
\begin{widetext}
\begin{equation}
\begin{aligned}
I_{l} (t) =&~
	\frac{4 \gamma_{l} \gamma_{r} C^2 \omega_0^2 }{\pi} \theta (t)
	\int d\varepsilon n^{l} (\varepsilon)
	\left|
		\frac
		{
			1
		}{
			(\varepsilon + i C (\gamma_{l} + \gamma_{r}))^2 - \overline{\omega}^2
		}
	\right|^2
\\&-
	\frac{4 \gamma_{l} \gamma_{r} C^2 \omega_0^2 }{\pi} \theta (t)
	\int d\varepsilon n^{r} (\varepsilon - V_{r})
\\&\times
	\left|
		-
		\frac
		{
			e^{-i \varepsilon t}
		}{
			(\varepsilon + i C (\gamma_{l} + \gamma_{r}))^2 - \overline{\omega}^2
		}
		-
		\frac{1}{2 \overline{\omega}}
		e^{- C ( \gamma_{l} + \gamma_{r}) t - i \overline{\omega} t}
		\left(
			- \frac{1}{ \varepsilon + i C ( \gamma_{l} + \gamma_{r}) - \overline{\omega} }
			+ \frac{1}{ \varepsilon - V_{r} + i C ( \gamma_{l} + \gamma_{r}) - \overline{\omega} }
		\right)
\right.\\&~~~+\left.
		\frac{1}{2 \overline{\omega}}
		e^{- C ( \gamma_{l} + \gamma_{r}) t + i \overline{\omega} t}
		\left(
			- \frac{1}{ \varepsilon + i C ( \gamma_{l} + \gamma_{r}) + \overline{\omega} }
			+ \frac{1}{ \varepsilon - V_{r} + i C ( \gamma_{l} + \gamma_{r}) + \overline{\omega} }
		\right)
	\right|^2.
\end{aligned}
\end{equation}
\end{widetext}
As we expected, if $t\rightarrow \infty$, then the current approaches the stationary value (\ref{cur_comp_93290}):
\begin{equation}
\begin{aligned}
I_{l} (t\rightarrow \infty) =&
	\frac{4 \gamma_{l} \gamma_{r} C^2 \omega_0^2}{\pi} \theta (t)
	\int d\varepsilon (n^{l} (\varepsilon) - n^{r} (\varepsilon - V_{r}) )
\\&\times
	\left|
		\frac
		{
			1
		}{
			(\varepsilon + i C (\gamma_{l} + \gamma_{r}))^2 - \overline{\omega}^2
		}
	\right|^2.
\end{aligned}
\end{equation}
If $t\rightarrow +0$, then there is no current at the other end of the chain, which illustrates the continuousness of the current value at $t = 0$:
\begin{equation}
\begin{aligned}
I_{l} (t\rightarrow +0) =&~
	\frac{4 \gamma_{l} \gamma_{r} C^2 \omega_0^2}{\pi} \theta (t)
	\int d\varepsilon (n^{l} (\varepsilon) - n^{r} (\varepsilon) )
\\&\times
	\left|
		\frac
		{
			1
		}{
			(\varepsilon + i C (\gamma_{l} + \gamma_{r}))^2 - \overline{\omega}^2
		}
	\right|^2
=0.
\end{aligned}
\end{equation}

If now, as in the previous section, we are interested in the role of "Majorana" states, we apply additional voltage to the right lead which is greater than the width of the localized states but less than the superconducting gap value.
That means that for $\varepsilon \lesssim \gamma_{l}, \gamma_{r}$ conditions $n^{l} (\varepsilon) = n^{r} (\varepsilon) = 0$, $n^{r} (\varepsilon -V_{r} ) =1$ hold.
The current  is determined by:
\begin{equation}
           \label{I(t)}
\begin{aligned}
I_{l} (t) =&
	- \frac{2 \gamma_{l} \gamma_{r} C \omega_0^2}{ \gamma_{l} + \gamma_{r} }
	\frac
	{
		1
	}{
		 \omega_{0}^2 + 4 C^2 \gamma_{l} \gamma_{r}
	}
	\theta (t)
\\&+
	\frac{ 2 \gamma_{l} \gamma_{r} C \omega_0^2}{ ( \gamma_{l} + \gamma_{r}) \overline{\omega}^2}
	e^{ - 2 C ( \gamma_{l} + \gamma_{r}) t} \theta (t)
\\&-
	i \gamma_{l} \gamma_{r} \frac{C^2 \omega_0^2}{ \overline{\omega}^2 }
	\frac
		{ e^{- 2 C ( \gamma_{l} + \gamma_{r}) t + 2 i \overline{\omega} t } \theta (t) }
		{ \overline{\omega} + i C ( \gamma_{l} + \gamma_{r}) }
\\&+
	i \gamma_{l} \gamma_{r} \frac{C^2 \omega_0^2}{ \overline{\omega}^2 }
	\frac
		{ e^{- 2 C ( \gamma_{l} + \gamma_{r}) t - 2 i \overline{\omega} t } \theta (t) }
		{ \overline{\omega} - i C ( \gamma_{l} + \gamma_{r}) }.
\end{aligned}
\end{equation}
We consider the case $\gamma_r,\gamma_l\gg \omega_0$ supposing that $\omega_0$ is always small. But for very symmetric tunneling coupling to the leads, we could have $\omega_0^2 \gg (\gamma_r-\gamma_l)^2$. This case looks unrealistic, nevertheless, it demonstrates an oscillating current signal at the left edge:
\begin{equation}
\begin{aligned}
I_{l} (t) =&
	- \frac{\omega_0^2}{2C( \gamma_{l} + \gamma_{r}) }
	\left[1-e^{- 2 C ( \gamma_{l} + \gamma_{r}) t} \right]
\\&-
	\left[
		\frac{ \omega_0}{ 2}\sin (2\omega_0t) -\frac{C( \gamma_{l} + \gamma_{r})}{2}
		(1-\cos(2\omega_0t))
	\right]
\\&~~\times
	e^{- 2 C ( \gamma_{l} + \gamma_{r}) t }.
\end{aligned}
\end{equation}

If $\omega_0 \ll |\gamma_r-\gamma_l|$ and $t>0$, equation (\ref{I(t)}) takes simpler form
\begin{equation}
           \label{I2(t)}
\begin{aligned}
I_{l} (t) = &
	- \frac{\omega_0^2}{2C( \gamma_{l} + \gamma_{r}) }
\\&\times
	\left[
		1 +
		\frac{ 4 \gamma_{l} \gamma_{r} }{ ( \gamma_{l} - \gamma_{r})^2} e^{- 2 C ( \gamma_{l} + \gamma_{r}) t }
\right.\\&~~~~~\left.
		-
		\frac{( \gamma_{l} + \gamma_{r})}{( \gamma_{l} - \gamma_{r})^2 }
		\left(\gamma_{l}e^{- 4 C  \gamma_{r}t}+\gamma_{r}e^{- 4 C  \gamma_{l}t}\right)
	\right].
\end{aligned}
\end{equation}
(Note that the minus sign corresponds to a current flowing from $r$ to $l$). For completely different tunneling rates, say $\gamma_r \gg \gamma_l$, the time evolution of the leading contribution to the current is determined by the slower  rate:
\begin{equation}
           \label{I3(t)}
I_{l} (t) =
	-\frac{\omega_0^2}{2C\gamma_{r} }\left[
	1-
	 e^{- 4 C  \gamma_{l}  t }
\right].
\end{equation}

The last formula shows, that if $\gamma_l \to 0$ the signal at the other end is very slow and very small.

\section{Nonstationary charge transfer}

Let us look at the last case  $\gamma_l \to 0$, which means $\gamma_l \ll \omega_0\ll \gamma_r$, from another point of view.
 Now we are interested not in the current but in changing the occupancy of the localized gap states at the left edge of the chain after the voltage change at the right end. So the applied voltage is supposed to be greater than $\gamma_r>\omega_0$ but less than the gap value the same as in the previous sections.
In our approach, the external reservoir is connected to the site $N$ and the occupancy of the states at site $1$ is measured. To investigate this occupancy of the states at site $1$, we need to calculate the time dependency of the Green function $G^{<}_{1, 1} (t, t)= i\, n_1(t) $. In the matrix representation, it is determined by the element $\{11\}$ of the matrix function $\widetilde \Gamma^{<}$,
\begin{equation}
   \label{Gamma<11}
\begin{aligned}
&\widetilde \Gamma^{<}_{1, 1} (t, t) =
	\int \frac{d\Omega}{2\pi} \frac{d\Omega'}{2\pi} e^{-i(\Omega - \Omega') t}
	\widetilde \Gamma^{R}_{1, N} (\Omega)
	\Sigma^{<}_{r} (\Omega, \Omega')
	\widetilde \Gamma^{A}_{N, 1} (\Omega'),
\end{aligned}
\end{equation}
where functions $\widetilde \Gamma^{R}_{1, N} (\Omega), \widetilde \Gamma^{A}_{N, 1} (\Omega')$ and $\Sigma^{<}_{r} (\Omega, \Omega')$ have been calculated in the previous sections (\ref{Gamma1N},\ref{Sigma<}) and in which we have set $\gamma_l=0$.

Calculating the equation (\ref{Gamma<11}), we obtain the following time dependence of $n_1(t)$:
\begin{equation}
\begin{aligned}
n_{1} (t) = &~
	 C
	\left[ 1
	-
	\frac{   \omega_0^2}{ \overline{\omega}^2}
	e^{ - 2 C \gamma_{r} t}  \right.
\\&+
	i \gamma_{r} \frac{C \omega_0^2}{ 2\overline{\omega}^2 }
	\frac
		{ e^{- 2 C \gamma_{r} t + 2 i \overline{\omega} t }  }
		{ \overline{\omega} + i C \gamma_{r} }
\\&-
	\left.  i \gamma_{r} \frac{C \omega_0^2}{ 2\overline{\omega}^2 }
	\frac
		{ e^{- 2 C \gamma_{r} t - 2 i \overline{\omega} t }  }
		{ \overline{\omega} - i C \gamma_{r} } 
		\right] \theta (t),
\end{aligned}
\end{equation}
where $\overline{\omega}$ is defined in (\ref{line_omega}).

For the case $\omega_0\ll\gamma_r$, which is more realistic in our opinion, we approximately have
\begin{equation}
\overline{\omega} = i C \gamma_{r} \left( 1 - \frac{\omega_{0}^2}{2 C^2 \gamma_{r}^2} - \frac{\omega_{0}^4}{8 C^4 \gamma_{r}^4} \right).
\end{equation}

\begin{figure}
\includegraphics[width=0.4\textwidth]{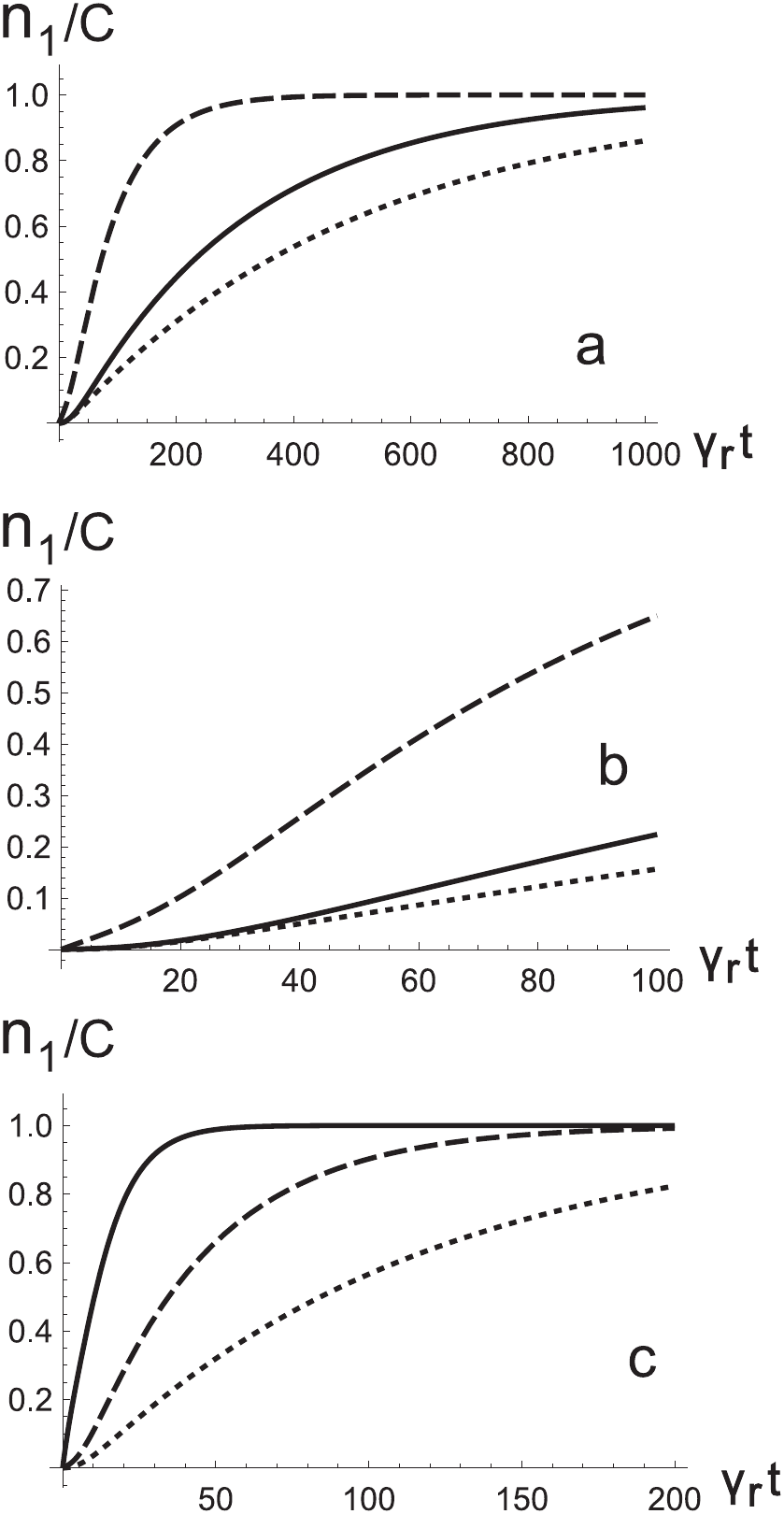}
\caption{Time dependence of the occupation number $n_1$ normalized to the equilibrium value $C$ for various parameters $C$ and $\omega_0/\gamma_{r}$. In panel (a) the parameters are: dashed line is for $C=0.03, \omega_0/\gamma_{r}=0.02$, solid line is for $C=0.03, \omega_0/\gamma_{r}=0.01$ and dotted line for $C=0.05, \omega_0/\gamma_{r}=0.01$.
Panel (b) shows on large scale the initial nonlinear section of the same time dependencies as in panel (a).
Panel (c) shows relaxation time growth with the increase of the chain length. Parameter $C=0.1$ for all curves. Solid line is for $\omega_0/\gamma_{r}=0.1$,  dashed line is for $\omega_0/\gamma_{r}=0.05$, and dotted line is for $\omega_0/\gamma_{r}=0.03$.
}
\label{graphs_sd}
\end{figure}

After the voltage change at the initial time moment $t=0$, we have the following filling dynamic at the other end of the chain:
\begin{equation}
\begin{aligned}
	\label{n_1}
n_{1} (t) =
	 C
	&\left[
		\left(1-e^{ - \displaystyle{\frac{\omega_{0}^2}{ C \gamma_{r} }} t } \right)
\right.\\&+\left.
		\frac
			{\omega_0^2}
			{ 4 C^2 \gamma_{r}^2 }\left(
		4e^{- 2 C \gamma_{r} t}
	  -
		 e^{- 4 C \gamma_{r} t }
		-
		3e^{- \displaystyle{\frac{\omega_{0}^2}{ C \gamma_{r} }} t } \right)
	\right]
\end{aligned}
\end{equation}

We see that only an exponentially small part of the full occupation, which is proportional to $\omega_0^2/\gamma_{r}^2 $, changes quickly, while relaxation to the new equilibrium occupation number $C$ is very slow.

The equilibrium occupation value $C$ is simply the value of the residue $\Gamma_{1,1}$ (\ref{state_454}) for the localized states at the first site. (This value
approaches $1/2$ in the highly symmetric case $\omega_0 \to 0, \Delta\to t$.)
The fast relaxation rate has the order of $\gamma_r$, while the slow rate is determined by the combination
$\gamma_{eff}=\omega_0^2/C\gamma_r$. Let us stress that the typical time of the slow relaxation exponentially grows with the increase of the length of the chain.

An example of large relaxation time is shown in Fig.\ref{graphs_sd}a. For this choice of parameters, the relaxation rate is three orders of magnitude smaller than the tunneling rate $\gamma_r$. The dependence (\ref{n_1}) for short time could be noticeably non-linear which is depicted in Fig.\ref{graphs_sd}b.
In Fig.\ref{graphs_sd}c it is shown how the relaxation time grows with decreasing $\omega_0$ (or with the chain length increasing).

Note that for the most interesting from the "topological protection" point of view case $\omega_0=0$, the value of $\gamma_{eff}=0$. This means that there is no possibility to transmit a signal to the other end of the chain.
So the notion that the localized edge states in the gap are  " the two halves of one Majorana state''  is wrong because  
if we change the occupation of one edge state the occupation of the other does not  change simultaneously.

\section{Conclusion}

The spectrum and transport properties of particle states in Kitaev chain can be investigated by the usual technique of single-particle Green's functions. For any non-stationary problem, this language is much more convenient than any other methods and allows to obtain exact analytical results.
It was demonstrated that the stationary tunneling current through the finite chain is always determined by the smallest of the rate parameters
$\omega_0^2/(\gamma_{l} + \gamma_{r}), \gamma_{l}, \gamma_{r}$  if the voltage bias is less than the superconducting gap.
For arbitrary parameters $\mu <|\Delta| <t$, the current is always exponentially small for long chains.
It should be noted that there can not be any noticeable peak at $\omega_0$ in the tunneling conductivity. And in the case $\omega_0=0$, the current vanishes completely.

The time-dependent behavior of the tunneling current after the sudden changing of bias voltage in one of the leads is also obtained. It was shown that typical time scales of tunneling current time evolution are determined mainly by the tunneling rates  $\gamma_{l}, \gamma_{r}$ from the left and right edge sites of the chain to the corresponding leads.
In the case of very weak tunneling coupling of one edge of the chain to the lead ($\gamma_{l}\ll \gamma_{r}$) we can investigate the characteristic time of charge transfer from the state at one end of the chain to the opposite edge state. We obtain that this time always exponentially increases with the growth of the chain length and the relaxation to the new equilibrium occupation number for the localized state is very slow.
Note that for the most interesting from the "topological protection" point of view case $\omega_0=0$ the value of $\gamma_{eff}=0$. This means that there is no possibility to transmit a signal to the other end of the chain.

So we see that any physical processes involving these ``Majorana'' localized states are either very slow or have very small amplitude. Thus, such states can hardly be used for any signal transfer, quantum information exchange and storage, and for quantum communications.

Let us add that similar behavior of non-stationary transmission is common for many systems with boundary states, and Kitaev chain has no advantage here despite its interesting topological properties.


\section*{Acknowlegements}

The authors thank for support grant RSF 23-22-00289.

\appendix
\section{Semi-infinite chain}
\label{appen}

Let us consider properties of a semi-infinite chain. If $N\rightarrow \infty$, equation (\ref{gamma_164}) takes the following form:
\begin{equation*}
\Gamma^{R}_{nm} (\omega) \rightarrow \Gamma^{(X)R}_{nm} (\omega),
\end{equation*}
where
\begin{equation}
\Gamma^{(X)R}_{nm} (\omega) \equiv
	\Gamma^{0R}_{nm} (\omega)
	-
	\Gamma^{0R}_{n0} (\omega)
	\left(
		 \Gamma^{0R}_{0,0} (\omega)
	\right)^{-1}
	\Gamma^{0R}_{0,m} (\omega).
\label{gamma_half_217}
\end{equation}
In turn, equation (\ref{det_gamma_164}) is replaced by
\begin{equation}
\det \left( \Gamma^{0R}_{0,0} (\omega) \right) = 0.
\label{det_gamma_half_923}
\end{equation}
Substituting values (\ref{gamma_values_749}) of the Green's functions into equation (\ref{det_gamma_half_923}) yields that $\omega$ should satisfy one of the two equations:
\begin{equation}
\begin{aligned}
&
	\frac
		{\mu + 2 t A_{+} + \omega}
		{\sqrt{A_{+}^2- 1} }
=
	\frac
		{\mu + 2 t A_{-} + \omega}
		{\sqrt{A_{-}^2- 1} },
\\&
	\frac
		{\mu + 2 t A_{+} - \omega}
		{\sqrt{A_{+}^2- 1} }
=
	\frac
		{\mu + 2 t A_{-} - \omega}
		{\sqrt{A_{-}^2- 1} }.
\label{eq_eq_eq}
\end{aligned}
\end{equation}
The direct substitution of $A_{+}$ and $A_{-}$ gives us
\begin{widetext}
\begin{equation}
     \label{R}
\mathcal{R}_{\pm} \equiv
	\frac
		{\mu + 2 t A_{\pm} + \omega}
		{\sqrt{A_{\pm}^2- 1} }
=
	\pm 2 \left|\Delta\right| \sqrt
	{
		\frac
		{
			\mu \left|\Delta\right|^2 \pm t \left|\Delta\right| \sqrt{\mu^2 + 4(\left|\Delta\right|^2 - t^2)\left(1 - \frac{\omega^2}{4\left|\Delta\right|^2}\right)} + \omega (\left|\Delta\right|^2 - t^2)
		}{
			\mu \left|\Delta\right|^2 \pm t \left|\Delta\right| \sqrt{\mu^2 + 4(\left|\Delta\right|^2 - t^2)\left(1 - \frac{\omega^2}{4\left|\Delta\right|^2}\right)} - \omega (\left|\Delta\right|^2 - t^2)
		}
	}.
\end{equation}
\end{widetext}
Here the square root in the rightmost part of the equation is defined in a way that its value approaches $+1$ when $\mu \rightarrow +\infty$. Here we should note that $\mathcal{R}_{-}$ as a function of $\mu$ has branch cuts at $\mu \in (2t - \left|\omega\right|; 2t + \left|\omega\right|)$, $\mu \in (- 2t - \left|\omega\right|; - 2t + \left|\omega\right|)$, which we need to account for to choose the correct value of the function. Here we assume that $\left|\omega\right|<2t$.

Let us note that from (\ref{R}) we obtain very usefull relation for $\omega=0$:
\begin{equation}
     \label{mu_A}
\frac
		{\mu + 2 t A_{\pm}}
		{\sqrt{A_{\pm}^2- 1} }=
\pm 2 \left|\Delta\right|
\end{equation}

Direct calculations show that both equations (\ref{eq_eq_eq}) have the same set of solutions. These solutions should satisfy
\begin{widetext}
\begin{equation}
\begin{aligned}
&
	\left(t \left|\Delta\right| \sqrt{\mu^2 + 4(\left|\Delta\right|^2 - t^2)\left(1 - \frac{\omega^2}{4\left|\Delta\right|^2}\right)} + \omega (\left|\Delta\right|^2 - t^2) \right)^2
=
	\left( t \left|\Delta\right| \sqrt{\mu^2 + 4(\left|\Delta\right|^2 - t^2)\left(1 - \frac{\omega^2}{4\left|\Delta\right|^2}\right)} - \omega (\left|\Delta\right|^2 - t^2) \right)^2.
\end{aligned}
\end{equation}
\end{widetext}
The two solutions that are given by
\begin{equation}
\mu^2 + 4(\left|\Delta\right|^2 - t^2)\left(1 - \frac{\omega^2}{4\left|\Delta\right|^2}\right) = 0
\end{equation}
do not actually solve equation (\ref{det_gamma_half_923}), since the nominator contains the same expression. It means that equation (\ref{det_gamma_half_923}) can only have a single solution at
\begin{equation*}
	\omega = 0.
\end{equation*}
The study of the properties of function $\mathcal{R}_{-}$ shows that this solution satisfies (\ref{eq_eq_eq}) if
\begin{equation}
\left|\mu\right| < 2t.
\end{equation}
Let us stress, that the solution $\omega = 0$   exists for any set of parameters under this condition. In particular, this solution does not require $\mu$ to be exactly zero.

To explore the properties of the bound state, we need to examine the behavior of Green's functions (\ref{gamma_half_217}) around the pole at $\omega \rightarrow 0$. We see that
\begin{widetext}
\begin{equation}
\begin{aligned}
\Gamma^{R}_{nm} (\omega \rightarrow 0) &=
	- \frac{1}{\omega + i\delta}
	\Gamma^{0R}_{n0} (0)
	\left(
		\left. \frac{\partial}{\partial \omega} \Gamma^{0R}_{0,0} (\omega) \right|_{\omega = 0}
	\right)^{-1}
	\Gamma^{0R}_{0,m} (0)
\\&=
	- \frac
	{
		\left|\Delta\right| (4 t^2 - \mu^2)
	}{
		2t ( 4 ( t^2 - \left|\Delta\right|^2 ) - \mu^2 )
	}
	\frac
	{
		(\chi_{+}^{\left|n\right|} - \chi_{-}^{\left|n\right|})
		(\chi_{+}^{\left|m\right|} - \chi_{-}^{\left|m\right|})
	}{
		\omega + i\delta
	}
	\begin{pmatrix}
		\theta (nm)
	&
		\frac{\Delta}{\left|\Delta\right|} \frac{\sign n + \sign m}{2}
	\\
		\frac{\Delta^{*}}{\left|\Delta\right|} \frac{\sign n + \sign m}{2}
	&
		\theta (nm)
	\end{pmatrix},
~~~~ \delta\rightarrow + 0,
\end{aligned}
\end{equation}
\end{widetext}
This function can be split into the sum of two functions, one of which is non-zero when both $n$ and $m$ are positive, and the other when both are negative. This represents the fact that, technically, the system described by Hamiltonian (\ref{ham_orig_short}) has two states at energy $\omega = 0$, one to the left, and the other to the right of the defect. If $n, m>0$, then
\begin{equation}
\begin{aligned}
\Gamma^{R}_{nm} (\omega \rightarrow 0) =&
	- \frac{1}{\omega + i\delta}
	\frac
	{
		\left|\Delta\right| (4 t^2 - \mu^2)
	}{
		2t ( 4 ( t^2 - \left|\Delta\right|^2 ) - \mu^2 )
	}
\\&\times
	(\chi_{+}^{n} - \chi_{-}^{n})
	(\chi_{+}^{m} - \chi_{-}^{m})
	\begin{pmatrix}
		1
	&
		\frac{\Delta}{\left|\Delta\right|}
	\\
		\frac{\Delta^{*}}{\left|\Delta\right|}
	&
		1
	\end{pmatrix}.
\end{aligned}
\label{pole_302}
\end{equation}
The way the branch cuts in $\chi_{+}$ and $\chi_{-}$ as functions of $\omega$ are defined leads to condition
\begin{equation*}
\left|\chi_{\pm}\right| \le 1.
\end{equation*}
for all values of the parameters.
If $\omega = 0$, $\left|\mu\right|<2t$, we can see that
\begin{equation}
\chi_{\pm} =
\frac
{
	- \mu
	\pm
	\sqrt{ \mu^2  - 4 (t^2 - \left|\Delta\right|^2) }
}{
	2 (t + \left|\Delta\right| )
}.
\label{chi_zero}
\end{equation}
For these values of parameters even stricter condition $\left|\chi_{\pm}\right| < 1$ applies. It means that the absolute values of the elements of matrix $\Gamma^{R}_{nm} (\omega \rightarrow 0)$ decrease exponentially for the chain elements toward the bulk of the chain.

Let us notice that coefficients (\ref{chi_zero}), which we have calculated without using ''Majorana formalism'', are identical to the values $X_{\pm}$ defined in \cite{kit}. This shows that our methods allow us to get the same results more straightforwardly and effectively.

In the highly symmetrical case $\mu = 0$ and $\left|\Delta\right| \rightarrow t$, we have
\begin{gather}
\begin{aligned}
\Gamma^{R}_{nm} (\omega \rightarrow 0) =&
	- \frac{1}{\omega + i\delta}
	\frac
	{
		t^2
	}{
		2 ( t^2 - \left|\Delta\right|^2 )
	}
\\&\times
	(\chi_{+}^{n} - \chi_{-}^{n})
	(\chi_{+}^{m} - \chi_{-}^{m})
	\begin{pmatrix}
		1
	&
		\frac{\Delta}{\left|\Delta\right|}
	\\
		\frac{\Delta^{*}}{\left|\Delta\right|}
	&
		1
	\end{pmatrix},
\end{aligned}
\\
\chi_{\pm} =
\pm
\sqrt{
	\frac
		{\left|\Delta\right| - t}
		{\left|\Delta\right| + t}
},
\end{gather}
which means that $\Gamma^{R}_{nm} (\omega \rightarrow 0) = 0$ for all $n$ and $m$ except for $n=m=1$. More precisely,
\begin{equation}
\Gamma^{R}_{nm} (\omega \rightarrow 0) =
	\frac{1}{2(\omega + i\delta)}
	\begin{pmatrix}
		1
	&
		\frac{\Delta}{\left|\Delta\right|}
	\\
		\frac{\Delta^{*}}{\left|\Delta\right|}
	&
		1
	\end{pmatrix}
	\delta_{n1} \delta_{m1}.
\end{equation}

Let us now check if the state given by (\ref{pole_302}) is normalized. To do this, we need to calculate the sum
\begin{equation}
\mathcal{N} = \sum_{n=1}^{+\infty} \left. \omega \Gamma^{R}_{nn} (\omega) \right|_{\omega\rightarrow 0}.
\end{equation}
Substituting (\ref{chi_zero}) into (\ref{pole_302}) and taking the sum over $n$, we get
\begin{equation}
\mathcal{N} =
	\frac{1}{2}
	\begin{pmatrix}
		1
	&
		\frac{\Delta}{\left|\Delta\right|}
	\\
		\frac{\Delta^{*}}{\left|\Delta\right|}
	&
		1
	\end{pmatrix}.
\end{equation}
Since $\Tr \mathcal{N} = 1$, the state is indeed normalized if we consider it in an abstract mathematical sense. However, the matrix's upper-left element is equal to $\frac{1}{2}$, which means that each such ''state'' contains only half of a particle.
 This suggests that, strictly speaking, this pole of the Green's function does not represent an actual state, but is an artifact of the ground state of the model. We refer to this point in the main text.

\end{document}